\documentclass[twocolumn,aps,prb,superscriptaddress,longbibliography]{revtex4-1}
\usepackage[colorlinks,bookmarks=false,citecolor=blue,linkcolor=red,urlcolor=blue]{hyperref}
\usepackage{verbatim}
\usepackage{color}
\usepackage{amsmath,amssymb,bm,braket,float,mathtools,subfigure}
\usepackage{graphicx,xfrac,appendix}
\usepackage[english]{babel}

\makeatletter

\usepackage{epstopdf}

\makeatother

\begin{document}

\title{Phase driven unconventional superradiance phase transition in non-Hermitian cascaded 
quantum Rabi cavities}

\author{Shujie Cheng}
\affiliation{Xingzhi College, Zhejiang Normal University, Lanxi 321100, China}
\affiliation{Department of Physics, Zhejiang Normal University, Jinhua 321004, China}

\author{Shuai-Peng Wang}
\affiliation{Beijing Academy of Quantum Information Sciences, Beijing 100193, China} 

\author{G. D. M. Neto}
 \email{gdmneto@zjnu.edu.cn}
\affiliation{Department of Physics, Zhejiang Normal University, Jinhua 321004, China}

\author{Gao Xianlong}
\email{gaoxl@zjnu.edu.cn}
\affiliation{Department of Physics, Zhejiang Normal University, Jinhua 321004, China}

\date{\today}

\begin{abstract} 
This study investigates phase-driven symmetry breaking leading to superradiance phase transitions in cascaded non-Hermitian quantum Rabi cavities. Non-Hermiticity is introduced via the phase coupling $\varphi$ between the atom and the optical field. In the thermodynamic limit of the quantum harmonic oscillator, we analytically derive the superradiance phase boundary, validated by observables. An unconventional quantum phase transition without a Hermitian analogue arises when $|\varphi|=\frac{\pi}{4}$ or $|\varphi|=\frac{3\pi}{4}$, where the phase boundary is uniquely determined by the cavity coupling, at $\mathcal{J}=\frac{1}{2}$, independent of the atom-photon coupling strength $g$. For other $\varphi$, the phase boundary relies on both $\mathcal{J}$ and $g$, similar to the scenario observed in Hermitian systems. Furthermore, we identify phase-driven first- and second-order superradiance phase transitions, focusing on the quantum criticality of the second-order transition by determining the critical exponents and the universality class. The feasibility of experimental realization is also discussed, aiming to inspire further studies on non-Hermitian superradiance quantum phase transitions.

\end{abstract}

\maketitle

\section{Introduction}

The cornerstone model for light-matter interaction is the quantum Rabi model (QRM), which involves a two-level atom (qubit) coupled through dipole interaction to a quantized single-mode cavity field and can be seen as the minimal version of the Dicke model\cite{Rabi, book, Dicke}. In recent years, there have been efforts to find exact analytical solutions for the QRM \cite{am_1, am_2, am_3, am_4, am_5, am_6, am_7, am_8}. Using the Bargmann space representation \cite{bs}, Braak obtained the transcendental function, known as the G-function, and discovered that the poles of the G-function were the exact solutions of the QRM \cite{es1}. Shortly after, Chen et al. achieved the exact solutions of a two-photon QRM by combining the Bogoliubov transformation and the G-function \cite{es2}. Over the past decade, these methods have been employed to calculate the exact solutions of generalized QRMs \cite{ext_1, ext_2, ext_3, ext_4, ext_5, ext_6, ext_7, ext_8, ext_9, ext_10, ext_11}. It has been found that symmetry breaking can lead to topological transitions in both the QRM \cite{tp1} and generalized QRM \cite{tp2}.

Parallel to the search for exact solutions, significant interest has emerged in exploring the quantum criticality of QRMs and generalized QRMs \cite{HP, ps1, ps2, R_d1, qc_1, qc_2, qc_3, zhangyy_1, zhangyy_2}. Quantum criticality denotes abrupt changes in a system's properties at quantum phase transitions (QPT), where macroscopic observables display non-analytic behavior due to quantum fluctuations. In the context of light-matter interaction, this phenomenon is exemplified by quantum phase transitions such as the superradiance phase transition (SPT), characterized by macroscopic excitations in ground-state populations\cite{review}.

Experimental observations in ion traps \cite{experiment} have confirmed a second-order QPT in the QRM, where researchers noted a rapid increase in both phonon number and internal ion excitations at the critical point. In the deep strong coupling regime of light-matter interaction models, a notable feature is the zero-temperature photon vacuum instability driving systems to undergo a QPT at critical coupling strength \( g_c \). Below \( g_c \), the system resides in a normal symmetric phase with the field in vacuum and atoms in their lowest electronic state. Above \( g_c \), both the field and atoms become excited, indicating a superradiant phase with broken symmetry. The thermodynamic limit is approached by increasing the number of particles to infinity in the Dicke model, resembling a classical limit for the atomic part. In contrast, phase transitions in single-atom systems occur at the classical oscillator limit obtained by increasing the qubit/oscillator frequency ratio.

Furthermore, the anisotropic quantum Rabi-Stark model (AQRSM), incorporating Stark coupling terms, demonstrates first-order and continuous QPTs \cite{xie2020first}. Notably, excited-state quantum phase transitions (ESQPTs) have been extensively studied for both the standard quantum Rabi model (QRM) \cite{puebla2016excited} and the AQRSM \cite{hu2023excited} in the limit of infinite frequency ratio. These transitions involve non-analytic changes in the excited-state properties as a control parameter varies, occurring at finite energy densities and marked by singularities in the density of states.
Studies have examined the universal dynamics and critical behaviors of the QRM, determining the exact phase boundary of the SPT using an effective low-energy approximation \cite{HP}. The critical exponent of the SPT, $\nu$, was found to be $\nu = \frac{3}{2}$ \cite{HP}. Analyzing a general scaling function has demonstrated that $\nu$ in the anisotropic QRM also satisfies $\nu = \frac{3}{2}$, and the phase boundary of the SPT was obtained via a semi-classical method \cite{ps2}. Subsequently, the exact $\nu$ was numerically confirmed by analyzing the critical behaviors of the finite frequency fidelity susceptibility \cite{qc_1}. Further research has revealed that for a multi-cavity system with two \cite{qc_2, qc_3} or three \cite{zhangyy_2} cascaded QRMs, the critical exponent remains the same as that of a single Rabi model. However, for a three-level QRM, a new critical exponent with $\nu = 1$ emerges \cite{zhangyy_3}. Notably, SPTs in QRMs depend on the atom-photon coupling strength, motivating investigation into the existence of unconventional SPTs independent of this coupling.

Recently, the combination of non-Hermiticity and QRMs has sparked intriguing developments. A generalized non-Hermitian integrable QRM was constructed, and its exact energy spectrum and eigenstates were obtained using the Bethe ansatz \cite{es_NH_1}. This work provides a generalized strategy to engineer integrable non-Hermitian spin-boson models. By introducing parity-time ($\mathcal{PT}$) symmetric gain and loss in the semi-classical Rabi model, it was discovered that $\mathcal{PT}$ symmetry breaking, accompanied by a real-complex transition in energy, occurs \cite{PT_Rabi_1, PT, PT1}. With the exact solutions of the $\mathcal{PT}$-symmetric Rabi models, the $\mathcal{PT}$-symmetry breaking point, i.e., the exceptional point (EP), can be located \cite{EP}. The Floquet $\mathcal{PT}$-symmetric semi-classical Rabi model was investigated, and under multiple-photon resonance conditions, the positions of the EPs from the exact Floquet spectrum were determined \cite{es_NH_2}. Using the adiabatic approximation, the exact energy spectrum of the $\mathcal{PT}$-symmetric QRM was obtained, locating the EPs \cite{es_NH_3}. With the emergence of EPs, observables present discontinuous behaviors. The exact energy spectra for one-photon and two-photon $\mathcal{PT}$-symmetric QRMs were solved using the Bogoliubov transformation and G-function, and the zeros of the G-function located the EPs \cite{es_NH_4}. Motivated by these studies, we will examine whether a fully real energy spectrum can emerge in a generalized non-Hermitian QRM without $\mathcal{PT}$ symmetry. Furthermore, this paper explores the quantum criticality of the model, revealing unique behaviors and transitions that distinguish it from conventional systems. By comparing and contrasting these findings with the established features of Hermitian systems, we aim to illuminate the intriguing parallels and divergences between the two realms.

The paper is organized as follows. In Sec.~\ref{S2}, we introduce the generalized non-Hermitian QRM without $\mathcal{PT}$ symmetry. In Sec.~\ref{S3}, we obtain the exact phase boundary of the model. In Sec.~\ref{S4}, we study the observables which signal the SPT. In Sec.~\ref{S5}, we discuss the quantum criticality of the SPT. Experimental feasibility is discussed in Sec.~\ref{S6}, and a summary is presented in Sec.~\ref{S7}.

\section{Two coupled non-Hermtian Rabi cavities}\label{S2} 
The generalized non-Hermitian QRM without $\mathcal{PT}$ symmetry is constructed by two cascaded non-Hermitian Rabi cavities, whose Hamiltonian ($\hbar=1$) is given as 
\begin{equation}\label{Ham}
H=\sum_{j=1,2}h_{j}+J(a^{\dag}_{1}+a_{1})(a^{\dag}_{2}+a_{2}),
\end{equation}
where $h_{j}$ describes the non-Hermitian Rabi cavity 
\[
h_{j}=\omega_{p}a^{\dag}_{j}a_{j}+\frac{\omega_{a}}{2}\sigma^{j}_{z}-\lambda e^{i\varphi}\left(a_{j}+a^{\dag}_{j}\right)\sigma^{j}_{x},
\]
with $j$ denoting the cavity index and $a^{\dag}_{j}$ ($a_{j}$) representing the creation (annihilation) operators of photons. Here, $\omega_{p}$ is the optical field frequency, $\omega_{a}$ is the atomic frequency, and $\lambda e^{i\varphi}$ is the coupling strength between the atom and optical field, where $\lambda$ represents the strength and $\varphi$ is the coupling phase. The non-Hermiticity of the system imposes that $\varphi \in (-\pi,0) \cup (0,\pi)$.

\begin{figure}[htp]
		\centering
		\includegraphics[width=0.5\textwidth]{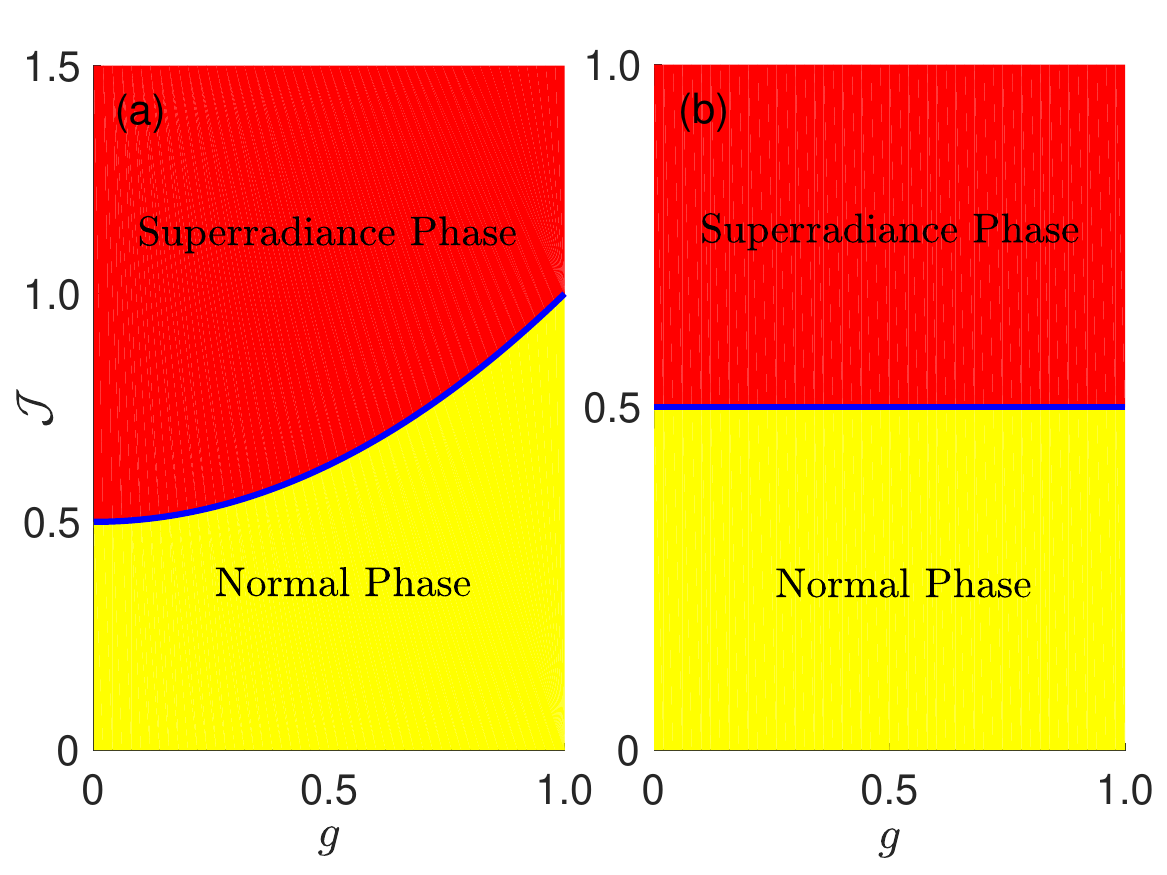}
		\caption{(Color online) Ground-state phase diagram in the $\mathcal{J}$-$g$ parameter space. 
		(a) $\varphi=\frac{\pi}{2}$. (b) $\varphi=\frac{\pi}{4}$. The blue solid lines denote the phase boundary, 
		satisfying $1-g^2\left(\cos^2\varphi-\sin^2\varphi\right)= 2\mathcal{J}$.  
		}
		\label{f1}
\end{figure}

In the context of a single non-Hermitian Rabi cavity, parity-time ($\mathcal{PT}$) symmetry is absent unless the coupling phase attains the specific values $\varphi=\pm\frac{\pi}{2}$. The time-reversal operator $\mathcal{T}$ enacts a complex conjugation operation, yielding $\mathcal{T} a^{\dag}_{j} (a_{j}) \mathcal{T} = a^{\dag}_{j} (a_{j})$. Similarly, the parity operator $\mathcal{P}$ performs an inversion transformation defined by $\mathcal{P}\hat{x}_{j} \mathcal{P} = -\hat{x}_{j}$ and $\mathcal{P}\hat{p}_{j} \mathcal{P} = -\hat{p}_{j}$ \cite{PT}. Consequently, we have
\begin{equation}
\begin{aligned}
\mathcal{PT} h_{j} \mathcal{TP}&=\omega_{p}a^{\dag}_{j}a_{j}-\mathcal{P}\left[\lambda e^{-i\varphi}\left(a_{j}+a^{\dag}_{j}\right)\sigma^{j}_{x}\right]\mathcal{P}+\frac{\omega_{a}}{2}\sigma^{j}_{z} \\
&=\omega_{p}a^{\dag}_{j}a_{j}+\lambda e^{-i\varphi}\left(a_{j}+a^{\dag}_{j}\right)\sigma^{j}_{x}+\frac{\omega_{a}}{2}\sigma^{j}_{z}. 
\end{aligned}
\end{equation}
Thus, the single non-Hermitian Rabi cavity manifests $\mathcal{PT}$-symmetry exclusively for $\varphi=\pm\frac{\pi}{2}$ (refer to an alternate proof in Ref. \cite{es_NH_4}). For any other values of $\varphi$, $\mathcal{PT}$-symmetry is not exhibited.

For the cascaded non-Hermitian QRM, we verify that this system is entirely not $\mathcal{PT}$-symmetric. The parity operator $\mathcal{P}$ mirrors operations between the two cavities, specifically $\mathcal{P}a^{\dag}_{j} (a_{j})\mathcal{P}=a^{\dag}_{j'} (a_{j'})$ and $\mathcal{P}\sigma^{j}_{z} (\sigma^{j}_{x})\mathcal{P}=\sigma^{j'}_{z} (\sigma^{j'}_{x})$ (where $j\neq j'$), while the time-reversal operator $\mathcal{T}$ satisfies $\mathcal{T}i\mathcal{T}=-i$. Performing $\mathcal{PT}$ on the cavity coupling term, we have 
\begin{equation}
\mathcal{PT} J\left(a^{\dag}_{1}+a_{1}\right)\left(a^{\dag}_{2}+a_{2}\right)\mathcal{TP}= J\left(a^{\dag}_{1}+a_{1}\right)\left(a^{\dag}_{2}+a_{2}\right).
\end{equation}
Performing $\mathcal{PT}$ on $h_{j}$, we find that
\begin{equation}
\begin{aligned}
\mathcal{PT} h_{j} \mathcal{TP} &= \omega_{p}a^{\dag}_{j'}a_{j'} - \lambda e^{-i\varphi}\left(a_{j'}+a^{\dag}_{j'}\right)\sigma^{j'}_{x} + \frac{\omega_{a}}{2}\sigma^{j'}_{z} \\
&\neq h_{j'},
\end{aligned}
\end{equation}
where $h_{j}$ is the Hamiltonian of the $j$-th cavity. Therefore, the cascaded non-Hermitian QRM does not exhibit $\mathcal{PT}$-symmetry.

\section{Analytical Results } \label{S3}
 To analytically solve the phase boundary 
of the ground-state superradiance phase transition, we introduce the dimensionless parameters $\eta=\frac{\omega_{a}}{\omega_{p}}$, $g=\frac{2\lambda}{\sqrt{\omega_{a}\omega_{p}}}$, 
and $\mathcal{J}=\frac{J}{\omega_{p}}$. Then, the Hamiltonian in Eq. (\ref{Ham}) can be rewritten as 
\begin{equation}\label{eq5}
\frac{H}{\omega_{p}}=\sum_{j=1,2}\frac{h_{j}}{\omega_{p}}+\mathcal{J}\left(a^{\dag}_{1}+a_{1}\right)\left(a^{\dag}_{2}+a_{2}\right), 
\end{equation}
in which
\begin{equation}
\frac{h_{j}}{\omega_{p}}=a^{\dag}_{j}a_{j}+\frac{\eta}{2}\sigma^{j}_{z}-\frac{g\sqrt{\eta}}{2}\left(a_{j}+a^{\dag}_{j}\right)\sigma^{j}_{x}.
\end{equation}
Performing the following transformation
\begin{equation}
\begin{aligned}
x_{j}&=\left(a^{\dag}_{j}+a_{j} \right)/\sqrt{2\eta}, \\
y_{j}&=i\left(a^{\dag}_{j}-a_{j}\right)\sqrt{\eta/2}, 
\end{aligned}
\end{equation}
and dividing both sides of Eq. (\ref{eq5}) by $\eta$, we obtain 
\begin{equation}
\mathcal{H}=\frac{H}{\omega_{p}\eta}\equiv\frac{1}{2}\sum_{j=1,2}\left(x^{2}_{j}+\frac{y^2_{j}}{\eta^2}+\sigma^{j}_{z}-\sqrt{2}g\sigma^{j}_{x}x_{j}\right)+2\mathcal{J}x_{1}x_{2}. 
\end{equation}

As studied in Refs.~\cite{ps1, ps2}, the superradiance phase transition (SPT) emerges under the infinite $\eta$ limit. Under this condition, we derive the analytical form of the Hamiltonian $\mathcal{H}$:
\begin{equation}
\mathcal{H} = \frac{1}{2} \sum_{j=1,2} \left[x_j^2 + \sigma_z^j - \sqrt{2}g \sigma_x^j x_j \right] + 2\mathcal{J} x_1 x_2.
\end{equation}
Diagonalizing $\mathcal{H}$, we obtain the energies $E_{\pm}(x_1, x_2)$:
\begin{equation}\label{analytic_E}
E_{\pm}(x_1, x_2) = \frac{1}{2} \sum_{j=1,2} \left(x_j^2 \pm \sqrt{1 + 2g^2 x_j^2} \right) + 2\mathcal{J} x_1 x_2.
\end{equation}
Expanding $E_{\pm}(x_1, x_2)$ to quadratic terms $\mathcal{O}(x_j^4)$, the low-energy branch $E_{-}(x_1, x_2)$ (representing the ground-state information) simplifies to:
\begin{equation}
E_{-}(x_1, x_2) = \frac{1}{2} \psi^T \Lambda \psi,
\end{equation}
where $\psi = (x_1, x_2)^T$ and $\Lambda$ is defined as:
\begin{equation}
\Lambda = \begin{pmatrix}
1 - g^2 e^{i2\varphi} & 2\mathcal{J} \\
2\mathcal{J} & 1 - g^2 e^{i2\varphi}
\end{pmatrix}.
\end{equation}

The matrix $\Lambda$ separates into its real part ${\rm Re}(\Lambda)$ and pure imaginary part ${\rm Im}(\Lambda)$, representing the background loss:
\begin{equation}\label{analytic_lambda}
\begin{aligned}
{\rm Re}(\Lambda) &= \left[1 - g^2 (\cos^2\varphi - \sin^2\varphi)\right] \sigma_0 + 2\mathcal{J} \sigma_x, \\
{\rm Im}(\Lambda) &= -i 2g^2 \cos\varphi \sin\varphi \sigma_0.
\end{aligned}
\end{equation}

\begin{figure}[htp]
		\centering
		\includegraphics[width=0.5\textwidth]{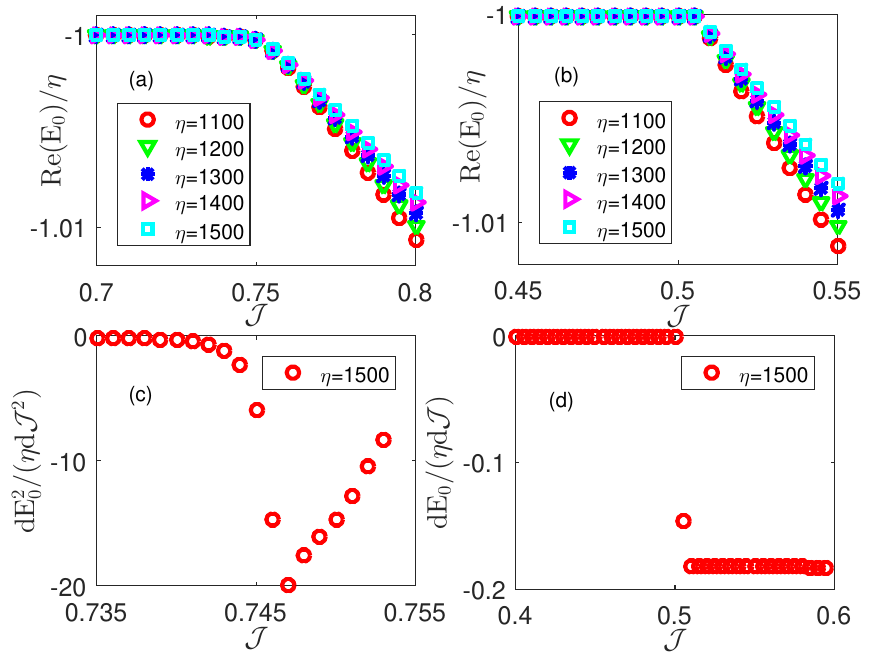}
		\caption{(Color online) The real part of the ground-state energy ${\rm Re}(E_{0})/\eta$ as 
		a function of $\mathcal{J}$ under different $\eta$ with $g=0.7$ and $\phi=\frac{\pi}{2}$ in (a) and 
		with $g=0.4$ and $\varphi=\frac{\pi}{4}$ in (b).  (c) The second-order derivative of ${\rm Re}(E_{0})/\eta$ 
		for $g=0.7$ and $\phi=\frac{\pi}{2}$. (d) The first-order derivative of ${\rm Re}(E_{0})/\eta$ 
		for $g=0.4$ and $\phi=\frac{\pi}{4}$. The photonic truncation number is $N=80$.  }
		\label{f2}
\end{figure}

The phase boundary of the superradiance phase transition (SPT) is primarily determined by ${\rm Re}(\Lambda)$ rather than ${\rm Im}(\Lambda)$. Diagonalizing ${\rm Re}(\Lambda)$ yields the eigenvalues $\epsilon_{\pm} = 1 - g^2 (\cos^2\varphi - \sin^2\varphi) \pm 2\mathcal{J}$. The condition for the SPT phase boundary, where the low-energy branch $\epsilon_{-} = 0$, results in $1 - g^2 (\cos^2\varphi - \sin^2\varphi) = 2\mathcal{J}$. This phase boundary reveals special phases at $\varphi = \pm\frac{\pi}{4}$ or $\varphi = \pm \frac{3\pi}{4}$, where $\mathcal{J} = \frac{1}{2}$ independently of the atom-photon coupling parameter $g$. Such a phase transition, independent of the coupling $g$, is termed \textit{unconventional} SPT as it has not been reported previously.

For other values of $\varphi$, the phase boundary depends jointly on $g$ and $\mathcal{J}$, categorizing it as conventional SPT. The subsequent analysis examines ground state energies, photon and qubit populations, fidelity, and fidelity susceptibility at $\varphi = \frac{\pi}{2}$ and $\varphi = \frac{\pi}{4}$ to characterize the properties of the non-Hermitian two-Rabi cavity system. Here, the ground state refers to the state with the lowest real part of the energy. Figures \ref{f1}(a) and \ref{f1}(b) illustrate phase diagrams for $\varphi = \frac{\pi}{2}$ and $\varphi = \frac{\pi}{4}$, respectively, where red regions indicate superradiance phases separated by blue curves representing phase boundaries, and normal phases are depicted in yellow regions.

\begin{figure}[htp]
		\centering
		\includegraphics[width=0.5\textwidth]{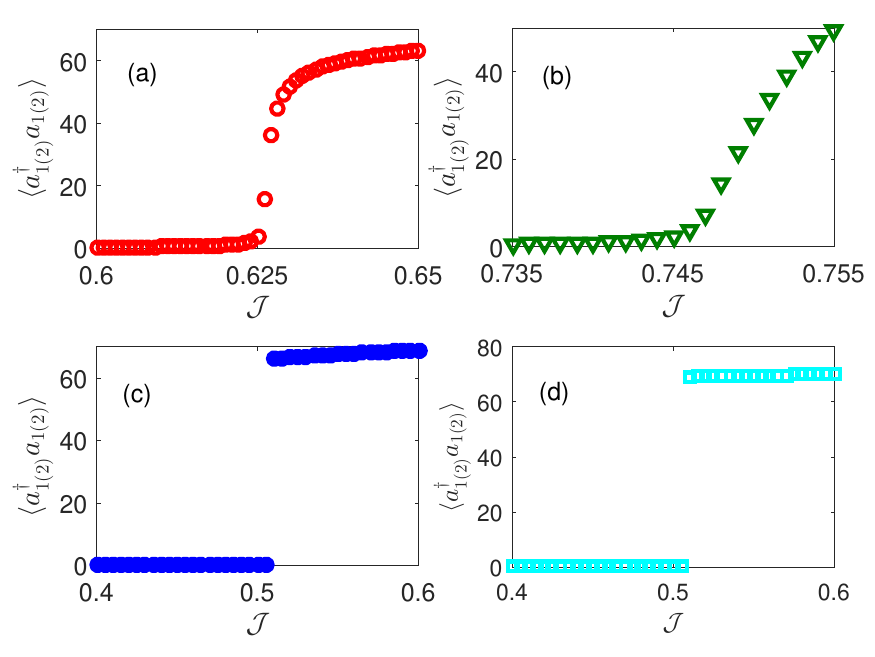}
		\caption{(Color online) The ground-state photon population $\langle a^{\dag}_{1(2)}a_{1(2)}\rangle$ as 
		a function of $\mathcal{J}$. (a) $g=0.5$ and $\phi=\frac{\pi}{2}$; 
		(b) $g=0.7$ and $\varphi=\frac{\pi}{2}$; (c) $g=0.4$ and $\varphi=\frac{\pi}{4}$; (d) $g=0.6$ and $\varphi=\frac{\pi}{4}$.  
		The photonic truncation number is $N=80$ and $\eta=1500$. \
		}
		\label{f3}
\end{figure}

\section{Signatures of the superradiance phase transition} \label{S4}
To identify the signatures and establish the phase boundary of the Superradiance Phase Transition (SPT), we commence our investigation by analyzing the ground-state energy. Energies are ranked based on their real parts, with the ground-state energy denoting the lowest real part.

Under the condition of $\varphi = \frac{\pi}{2}$ and varying $\eta$, the real parts of the ground-state energies $\mathrm{Re}(E_{0})/\eta$ as a function of $\mathcal{J}$ for $g=0.7$ are depicted in Fig.~\ref{f2}(a). Here, the photonic truncation number $N=80$ ensures the convergence of results. It is observed that with the onset of the SPT, $\mathrm{Re}(E_{0})/\eta$ for different $\eta$ decreases from $-1$ to lower values. According to the analytical predictions, the transition point under $\varphi = \frac{\pi}{2}$ and $g=0.7$ is expected to occur at $\mathcal{J}=0.745$, marked by the vertical black line. This confirms the occurrence of the SPT at the theoretically derived transition points.

Similarly, the plot in Fig.~\ref{f2}(b) illustrates the relationship between $\mathrm{Re}(E_{0})/\eta$ and $\mathcal{J}$ for $\varphi = \frac{\pi}{4}$ with $g=0.4$. As in the $\varphi = \frac{\pi}{2}$ case, when $\mathcal{J}$ surpasses the transition point, $\mathrm{Re}(E_{0})/\eta$ for different $\eta$ decreases from $-1$ to lower values, indicating the onset of the SPT. The analytical phase boundary indicates that the transition point under $\varphi = \frac{\pi}{4}$ should be at $\mathcal{J}=0.5$, which aligns with the observed transitions for $g=0.4$.

The analysis of Fig.~\ref{f1}(c) shows that the second-order derivative of $\mathrm{Re}(E_{0})/\eta$ under $\varphi = \frac{\pi}{2}$ and $g=0.7$ is discontinuous at the transition point, characteristic of a second-order quantum phase transition. Conversely, Fig.~\ref{f1}(d) demonstrates that the first-order derivative of $\mathrm{Re}(E_{0})/\eta$ under $\varphi = \frac{\pi}{4}$ and $g=0.4$. These findings indicate that the conventional SPT is a second-order quantum phase transition, while the unconventional SPT is a first-order quantum phase transition.

Specifically, cases where $|\varphi|=\frac{\pi}{4}$ and $|\varphi|=\frac{3\pi}{4}$ correspond to second-order quantum phase transitions, whereas other values of $\varphi$ correspond to first-order quantum phase transitions. However, this study focuses primarily on conventional and unconventional superradiance phase transitions, thus excluding detailed consideration of results for other $\varphi$ values.

 \begin{figure}[htp]
		\centering
		\includegraphics[width=0.5\textwidth]{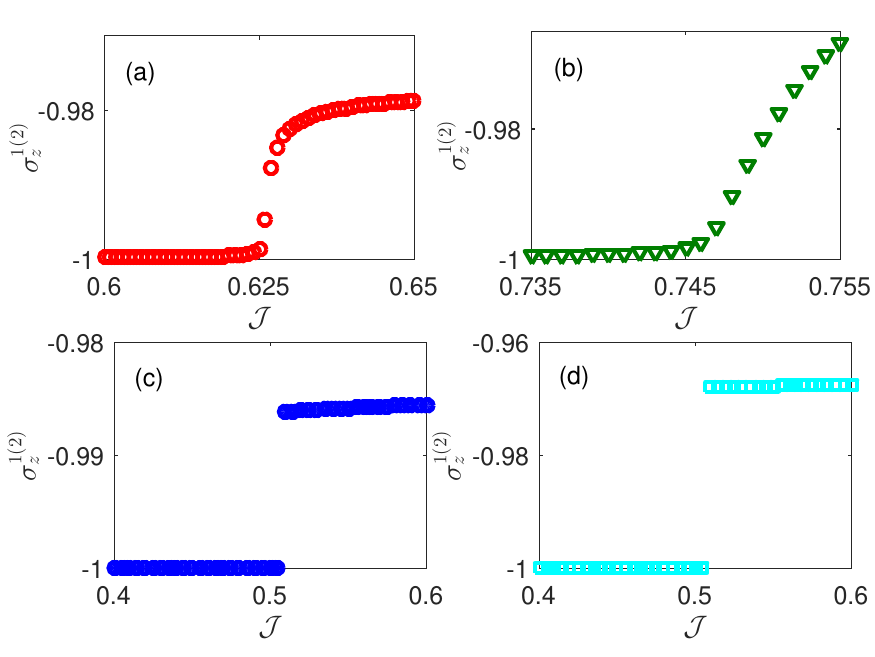}
		\caption{(Color online) The ground-state atom population $\sigma^{1(2)}_{z}$ as 
		a function of $\mathcal{J}$. (a) $g=0.5$ and $\phi=\frac{\pi}{2}$; 
		(b) $g=0.7$ and $\varphi=\frac{\pi}{2}$; (c) $g=0.4$ and $\varphi=\frac{\pi}{4}$; (d) $g=0.6$ and $\varphi=\frac{\pi}{4}$.  
		The photonic truncation number is $N=80$ and $\eta=1500$. 
		}
		\label{f4}
\end{figure}

Except for the ground-state energy, the phenomenon of SPT can be readily seen from 
the ground-state photon population. Under $\varphi=\frac{\pi}{2}$, ground-state photon populations $\langle a^{\dag}_{1(2)}a_{1(2)}\rangle$ 
for $g=0.5$ and $g=0.7$ are plotted in  Fig.~\ref{f3}(a) and \ref{f3}(b), respectively. It is seen that there is no photon 
excitation when $\mathcal{J}$ is smaller than the transition point, but there are macroscopic photon excitations when 
the cavity coupling parameter $\mathcal{J}$ exceeds the transition point. The phenomenon is similar to the $\varphi=\frac{\pi}{4}$ 
case. For $g=0.4$ and $g=0.6$, the corresponding ground-state photon populations are plotted in Fig.~\ref{f3}(c) and \ref{f3}(d), 
respectively. We can see that there are macroscopic photon excitations when $\mathcal{J}$ is larger than $0.5$, while the 
photon excitations are evidently suppressed when $J<0.5$. 

In the superradiance phase, in addition to the macroscopic excitations of the photons, the atoms also become excited and active. To verify this, we plot the populations of atoms $\langle \sigma^{1(2)}_{z} \rangle$ as a function of $\mathcal{J}$ for different $\varphi$ and $g$ in Fig.~\ref{f4}(a)-\ref{f4}(d). When $\mathcal{J}$ is below the transition point, the atoms remain in their ground state with populations $\langle \sigma^{1(2)}_{z} \rangle = -1$. However, as $\mathcal{J}$ crosses the transition point, the populations deviate from -1, indicating that the atoms are excited and active. These excited atoms drive the macroscopic photon excitations observed in the SPT.

While the photon excitations and atom populations of the system change noticeably with the occurrence of SPT, the two types of SPTs exhibit distinct observable behaviors. As illustrated in Fig.~\ref{f3}, during a conventional SPT, $\langle a^{\dag}_{1(2)}a_{1(2)} \rangle$ continuously varies to macroscopic values. In contrast, during an unconventional SPT, $\langle a^{\dag}_{1(2)}a_{1(2)} \rangle$ abruptly jumps to macroscopic values without an asymptotic process. This distinction is similarly evident in the atomic populations shown in Fig.~\ref{f4}.

 \begin{figure}[htp]
		\centering
		\includegraphics[width=0.5\textwidth]{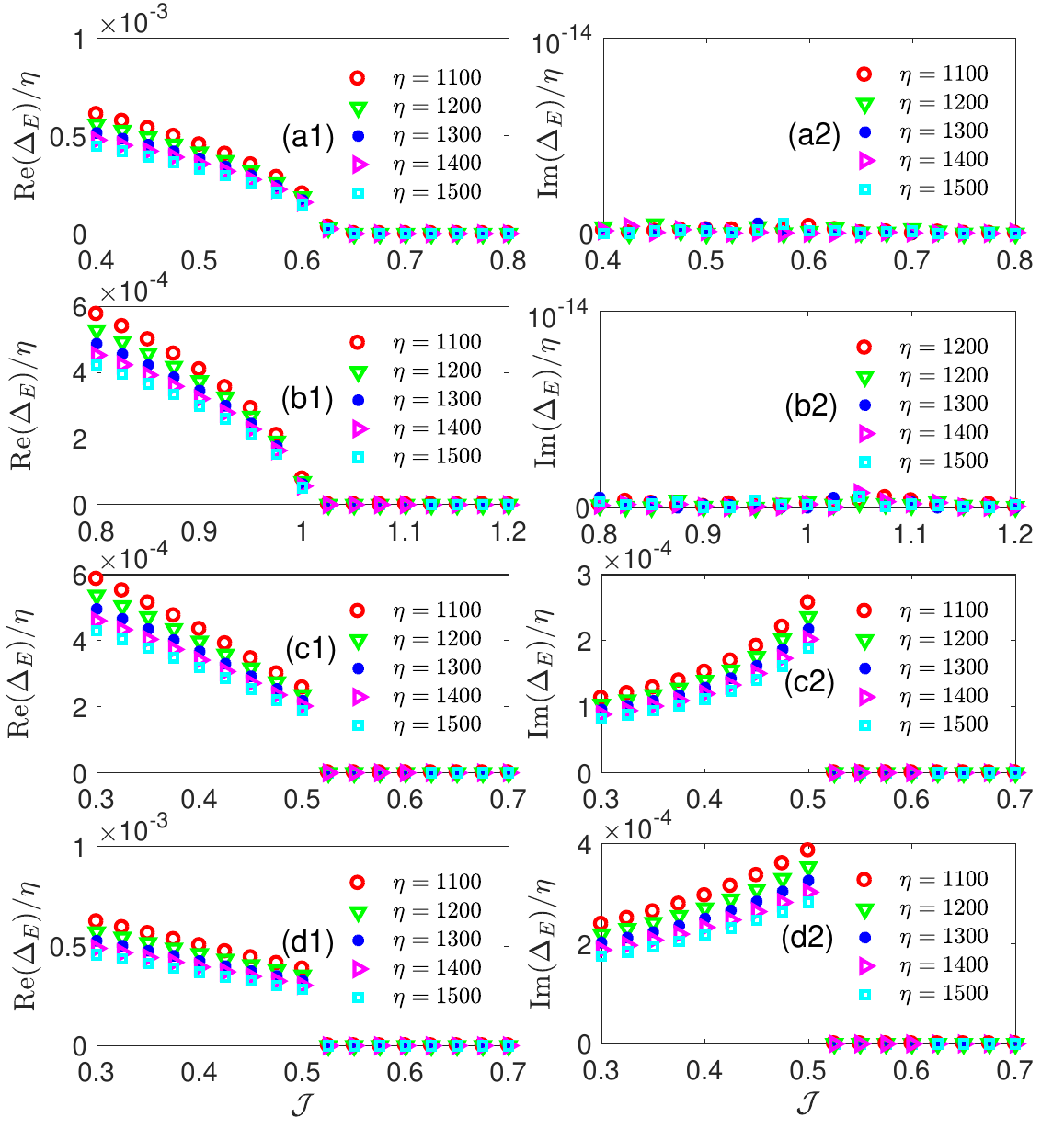}
		\caption{(Color online) The excitation gaps  ${\rm Re}(\Delta_{E})$ and ${\rm Im}(\Delta_{E})$ as 
		functions of $\mathcal{J}$ under different $\eta$. (a1) and (a2): $g=0.5$ and $\phi=\frac{\pi}{2}$; 
		(b1) and (b2): $g=1$ and $\varphi=\frac{\pi}{2}$; (c1) and (c2):  $g=0.4$ and $\varphi=\frac{\pi}{4}$; 
		(d1) and (d2): $g=0.6$ and $\varphi=\frac{\pi}{4}$. The photonic truncation number is $N=36$. 
		}
		\label{f5}
\end{figure}

In addition to the photon and atom populations, the excitation gaps associated with the two types of SPTs also exhibit distinct behaviors. These excitation gaps have two components: the real excitation gap ${\rm Re}(\Delta_{E})/\eta$, which is the gap between the real parts of the lowest two energies, and the imaginary excitation gap ${\rm Im}(\Delta_{E})/\eta$, which is the gap between the imaginary parts of the lowest two energies. Figures \ref{f5}(a1)-\ref{f5}(d1) show ${\rm Re}(\Delta_{E})/\eta$ as a function of $\mathcal{J}$ for different $\eta$, while Figures \ref{f5}(a2)-\ref{f5}(d2) present ${\rm Im}(\Delta_{E})/\eta$ under the same conditions.

Regardless of whether $\varphi = \frac{\pi}{2}$ or $\varphi = \frac{\pi}{4}$, the real excitation gap undergoes a closing process when $\mathcal{J}$ surpasses the transition point. However, a noticeable difference arises in the imaginary excitation gaps between the two cases. As seen in Figures \ref{f5}(a2) and \ref{f5}(b2), ${\rm Im}(\Delta_{E})/\eta$ remains zero (values much smaller than $10^{-14}$ can be numerically considered as zero) both before and after the transition point. In contrast, Figures \ref{f5}(c2) and \ref{f5}(d2) demonstrate that the imaginary excitation gaps experience a closing process similar to that of the real excitation gaps.

\begin{figure}[htp]
		\centering
		\includegraphics[width=0.5\textwidth]{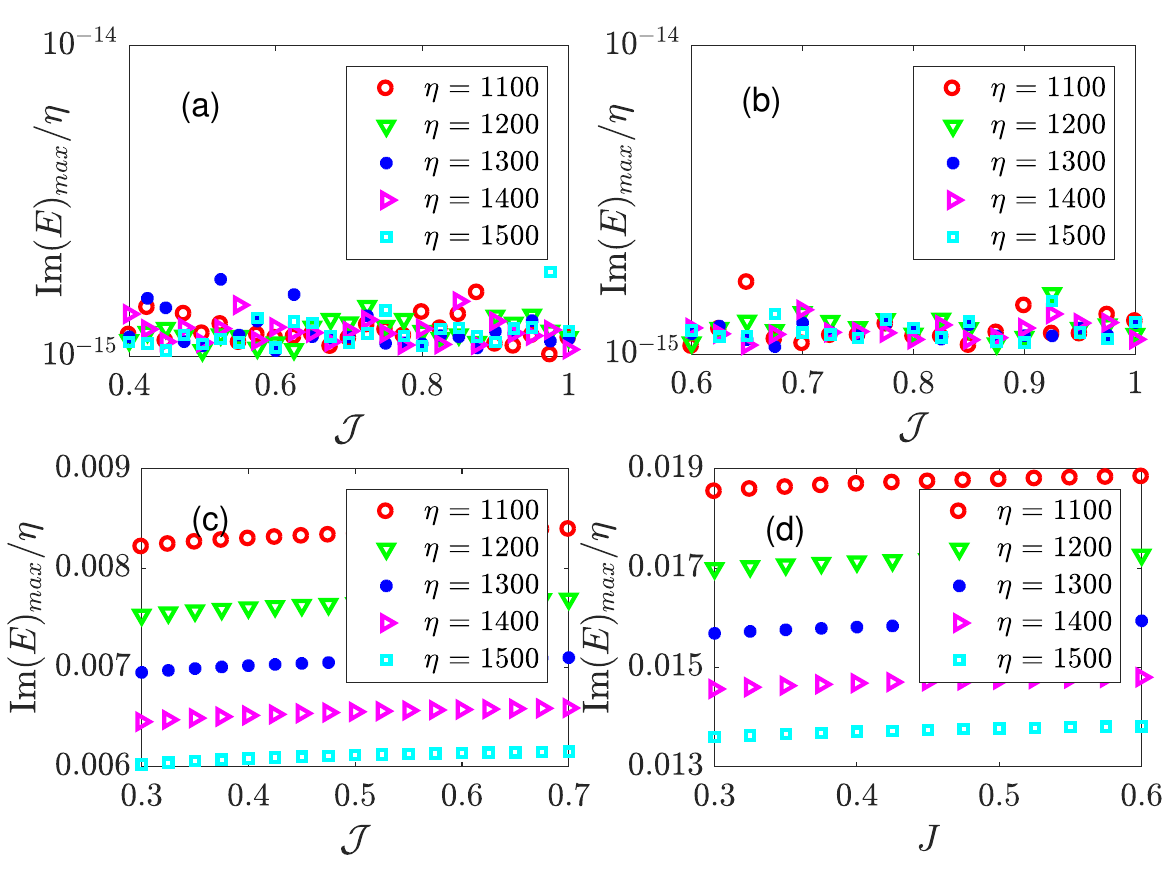}
		\caption{(Color online) The maximal imaginary part of the energy spectrum ${\rm Im}(E)_{max}/\eta$ as 
		functions of $\mathcal{J}$ under different $\eta$. (a) $\varphi=\frac{\pi}{2}$ and $g=0.7$. 
		(b) $\varphi=\frac{\pi}{2}$ and $g=0.8$. (c) $\varphi=\frac{\pi}{4}$ and $g=0.4$. 
		(d) $\varphi=\frac{\pi}{4}$ and $g=0.6$.   
		The photonic truncation number is $N=36$. 
		}
		\label{f6}
\end{figure}

For a $\mathcal{PT}$-symmetric Rabi system, a real-to-complex transition in the energy spectrum is well-established \cite{PT,PT1,es_NH_2,es_NH_3,es_NH_4}. Remarkably, the non-Hermitian two-Rabi cavity system without $\mathcal{PT}$ symmetry also exhibits a real energy spectrum. Under the condition $\varphi=\frac{\pi}{2}$, the maximum imaginary part of the energy spectra for $g=0.7$ and $g=0.8$, denoted as ${\rm Im}(E)_{\text{max}}$, is plotted in Figs.~\ref{f6}(a) and \ref{f6}(b), respectively. The analytically determined superradiance transition point is $\mathcal{J}=0.745$ for $g=0.7$ and $\mathcal{J}=0.82$ for $g=0.8$. ${\rm Im}(E)_{\text{max}}/\eta$ approaches zero both before and after the superradiance transition point, indicating that the energy spectra are real and no real-to-complex transition occurs.

In contrast, the spectral properties under $\varphi=\frac{\pi}{4}$ are distinct. Figures \ref{f6}(c) and \ref{f6}(d) show ${\rm Im}(E)_{\text{max}}/\eta$ for $g=0.4$ and $g=0.6$, respectively. Unlike the $\varphi=\frac{\pi}{2}$ case, the energy spectra under $\varphi=\frac{\pi}{4}$ are complex due to the non-zero ${\rm Im}(E)_{\text{max}}/\eta$. This distinction can be analytically understood from Eq. (\ref{analytic_E}) and Eq. (\ref{analytic_lambda}). When $\varphi=\frac{\pi}{2}$, the imaginary part vanishes, whereas for $\varphi=\frac{\pi}{4}$, it persists. The numerical results in Fig.~\ref{f6} confirm these analytical insights.

The full real nature of the energy spectra for $\varphi=\frac{\pi}{2}$ suggests that the conventional SPT in this case shares the universality class with Hermitian QRM systems \cite{HP,ps2,qc_1,qc_2,qc_3,zhangyy_1}, allowing for derivation of the critical exponent from its critical behavior. Conversely, for the complex energy spectrum case ($\varphi=\frac{\pi}{4}$), we hypothesize the absence of critical behavior. Although quantum criticality has been reflected in photon and atom populations for conventional SPT and its absence for unconventional SPT, we aim to validate these conjectures further through examination of ground-state fidelity and fidelity susceptibility.

\begin{figure}[htp]
		\centering
		\includegraphics[width=0.5\textwidth]{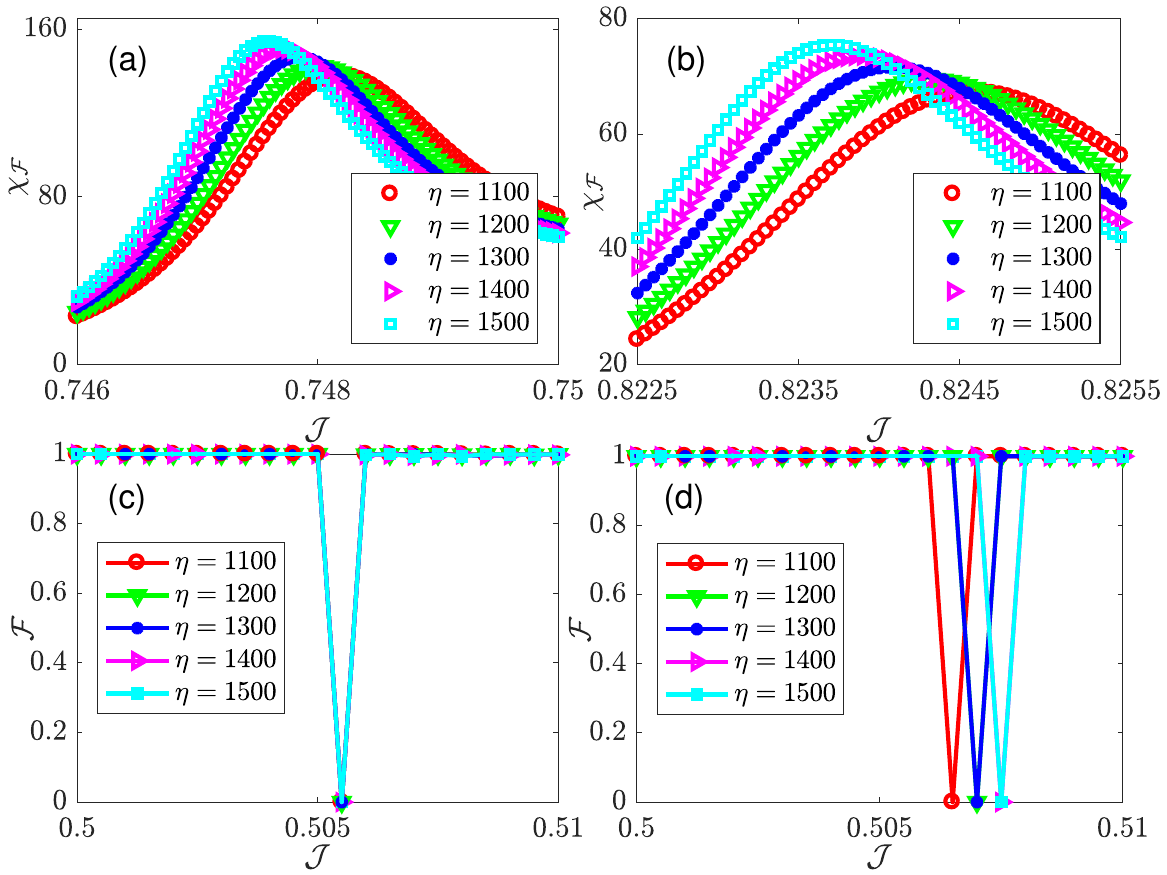}
		\caption{(Color online) Fidelity susceptibility $\chi_{\mathcal{F}}$ as functions of $\mathcal{J}$ under different $\eta$ 
		with $\varphi=\frac{\pi}{2}$ and $g=0.7$ in (a), and with $\varphi=\frac{\pi}{2}$ and $g=0.8$ in (b). 
		The involved parameter is $\delta\mathcal{J}=10^{-5}$. 
               Fidelity $\mathcal{F}$ as functions of $\mathcal{J}$ under different $\eta$ with $\varphi=\frac{\pi}{4}$ and $g=0.4$ in (c), 
               and with $\varphi=\frac{\pi}{4}$ and $g=0.6$ in (c). The involved parameter is $\delta\mathcal{J}=5\times 10^{-4}$. 
		The photonic truncation number in all calculations is $N=80$. 
		}
		\label{f7}
\end{figure}

\section{Comparative Study of Conventional vs. Unconventional Superradiance Phase Transitions} \label{S5}

We focus on the quantum phase transition induced by the cavity coupling parameter $\mathcal{J}$. The fidelity for the ground state $\mathcal{F}$ is defined as \cite{fidelity_0,fidelity_1,fidelity_2,fidelity_3}:
\begin{equation}
\mathcal{F}=|\langle \psi_{0}(\mathcal{J})|\psi_{0}(\mathcal{J}+\delta\mathcal{J})|\rangle|, 
\end{equation}
which measures the overlap of the ground state before and after a small change $\delta\mathcal{J}$ in $\mathcal{J}$. According to Refs. \cite{FS_1,FS_2,FS_3,FS_4,FS_5}, the fidelity susceptibility $\chi_{\mathcal{F}}$ is given by:
\begin{equation}
\chi_{\mathcal{F}}=\lim_{\delta \mathcal{J}\rightarrow 0}\frac{-2\ln \mathcal{F}(\mathcal{J},\delta \mathcal{J})}{(\delta \mathcal{J})^2}. 
\end{equation}
If a quantum phase transition exists, $\mathcal{F}$ exhibits a drop at the transition point, while $\chi_{\mathcal{F}}$ peaks.

\begin{figure}[htp]
		\centering
		\includegraphics[width=0.5\textwidth]{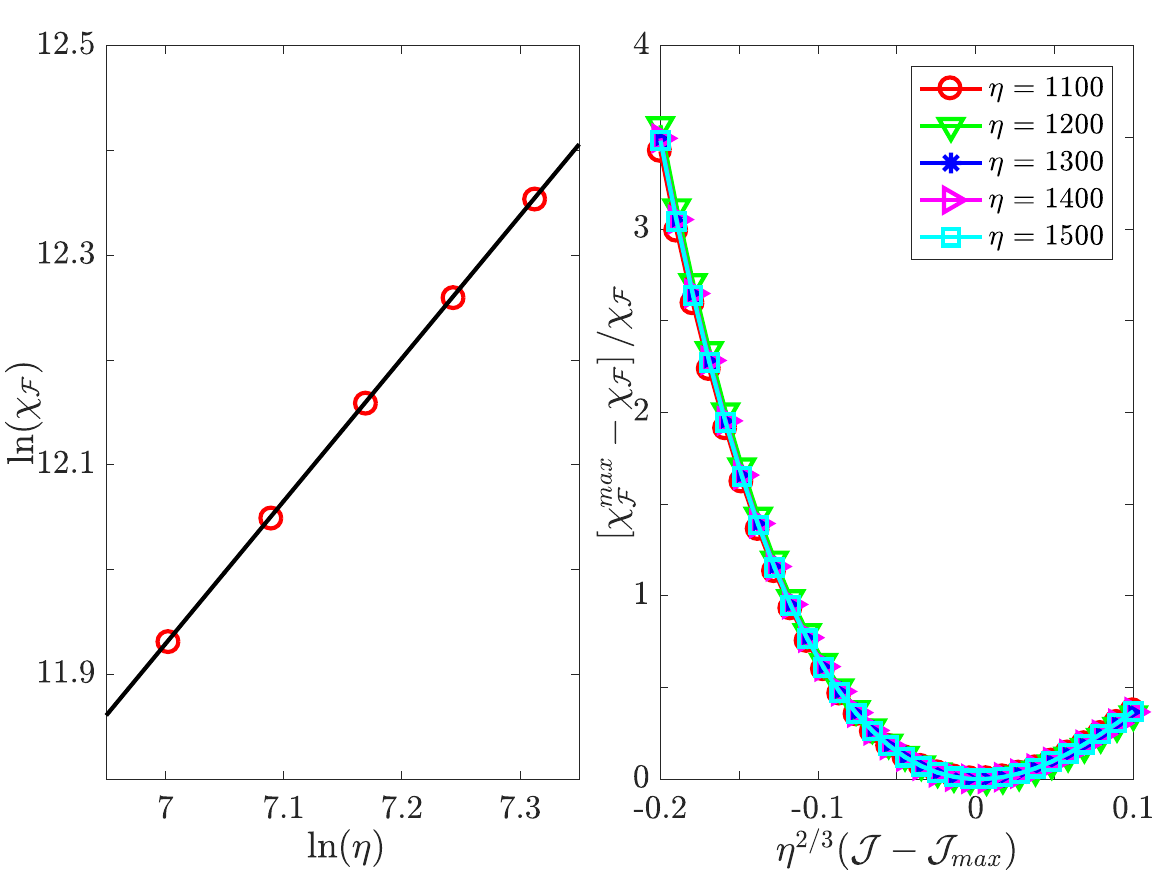}
		\caption{(Color online) (a) Universal finite $\eta$ scaling relation of the maximal fidelity susceptibility $\chi^{max}_{F}$. 
		(b) Data collapse of $\left[\chi^{max}_{\mathcal{F}}-\chi_{\mathcal{F}}\right]/\chi_{\mathcal{F}}$. 
		The involved parameters are $\varphi=\frac{\pi}{2}, $$g=0.7$, and $\delta J=10^{-5}$.	
				}
		\label{f8}
\end{figure}

Under $\varphi=\frac{\pi}{2}$, $\chi_{\mathcal{F}}/\eta$ for $g=0.7$ and $g=0.8$ are shown in Figs.~\ref{f7}(a) and \ref{f7}(b), respectively. As $\mathcal{J}$ increases, $\chi_{\mathcal{F}}/\eta$ smoothly reaches a peak and subsequently decreases. Notably, $\chi_{\mathcal{F}}/\eta$ displays clear critical behavior; as $\eta$ increases, the peak at the coupling parameter $\mathcal{J}_{max}$ approaches the exact solution. Conversely, for $\varphi=\frac{\pi}{4}$, no critical phenomena are observed. 
As seen in Figs.~\ref{f7}(c) and \ref{f7}(d), the fidelity curves sharply drop to zero at a single $\mathcal{J}$ value without any asymptotic behavior between the fidelity valley (or $\chi_{\mathcal{F}}$ peak) and $\eta$. This indicates the presence of quantum criticality in conventional SPT and its absence in unconventional SPT.

Next, we analyze the quantum criticality of conventional SPT ($\varphi=\frac{\pi}{2}$). From the $\chi_{\mathcal{F}}$ vs. $\eta$ curves, we extract the maximal fidelity susceptibility $\chi^{max}_{F}$ for a specific $\eta$. The universal finite $\eta$ scaling relation of $\chi_{\mathcal{F}}$ for $g=0.7$ is depicted in Fig.~\ref{f8}(a). The fitted black curve satisfies $\ln\chi_{\mathcal{F}}=(-1.363\pm 0.004) \ln \eta+2.388$, indicating a scaling exponent $\mu=-1.363\pm 0.004$. Following the relationship $\mu=\frac{2}{\nu}$ \cite{qc_1,RMP2011}, we derive $\nu=1.4674\pm0.004$. Although numerically slightly deviating from the exact solution ($\nu_{exact}=3/2$) of the single Rabi model \cite{HP,ps2}, we argue that the quantum phase transition under $\varphi=\frac{\pi}{2}$ shares the same universality class as the single Rabi model. Around the transition point, $\chi_{\mathcal{F}}$ scales as $\left[\chi^{max}_{\mathcal{F}}-\chi_{\mathcal{F}}\right]/\chi_{\mathcal{F}}=f\left[\eta^{1/\nu}\left(\mathcal{J}-\mathcal{J}_{max}\right)\right]$, where $f(x)$ is a universal scaling function. From Fig.~\ref{f8}(b), across various $\eta$, we observe that the data collapse into a single curve with the critical exponent $\nu=\frac{3}{2}$, indicating the same universality class as the single Rabi model \cite{HP,ps2}. Indeed, similar critical exponents ($\nu=3/2$) are observed in the Dicke model \cite{Dicke_1,Dicke_2} and the Lipkin-Meshkov-Glick model \cite{LMG_1,LMG_2,LMG_3}, further affirming that the conventional SPT in the non-Hermitian cascade Rabi cavities studied here belongs to this universal class.

\section{Experimental feasibility}\label{S6}

The experimental exploration of the phenomena discussed here could be facilitated by a proposed circuit-QED 
architecture, as detailed in Ref. \cite{SC}. This setup comprises two resonators, each equipped with an embedded 
artificial atom, specifically a superconducting flux qubit. The key feature of this design is the implementation of 
a coupling mechanism between the resonators that is both broadly and rapidly tunable. By leveraging fast-modulating 
fields, this tunable coupling can effectively emulate a phase-dependent interaction term, a crucial component for 
investigating PT-related physics. Furthermore, the architecture allows for the coupling between the flux qubits and 
the resonators to be pushed into the ultrastrong coupling regime ($g>0.1$), and even the deep-strong coupling 
regime ($g>1$). These advanced capabilities of the circuit-QED system provide a solid platform not only for testing 
the theoretical predictions presented in this article but also for realizing these concepts in practical, soon-to-be-available devices.

Similarly, experiments exploring our findings can also be conducted in ion trap setups. Ion traps utilize individual ions confined in harmonic traps as qubits, offering precise control over quantum states and interactions. This setup enables the emulation of phase-dependent interactions required to investigate PT-related physics, similar to circuit-QED architectures. The trapped-ion system offers a versatile platform combining laser-coupled spin and bosonic degrees of freedom, ideal for simulating Hamiltonians of light-matter interactions. Demonstrated capabilities include quantum simulations of many-body spin models \cite{experiment2}, single-ion quantum Rabi models \cite{experiment}, and recently, the Rabi-Hubbard (RH) model with up to 16 ions \cite{experiment3}, exploring equilibrium phase transitions and quantum dynamical properties using spin observables. The scalability and controllability of ion trap systems make them suitable for exploring both conventional and unconventional superradiance phase transitions in non-Hermitian quantum Rabi-dimer models and other related phenomena.

\section{Summary}\label{S7}
Exploring a non-Hermitian Quantum Rabi Model composed of two cascaded non-Hermitian Rabi cavities, we investigate its superradiance phase boundary akin to its Hermitian counterpart. As the system undergoes SPT, transitioning from a quantum vacuum state to one that becomes macroscopically populated, we observe unique exotic phenomena arising from non-Hermitian effects.

Notably, for specific atom-photon coupling phases such as $\varphi=\pm\frac{\pi}{4}$ or $\varphi=\pm\frac{3\pi}{4}$, the superradiance phase boundary remains constant, independent of the coupling strengths. This suggests an unconventional SPT. In contrast, for other coupling phases, the phase boundary depends on both the cavity and atom-photon coupling strengths, indicative of a conventional SPT.

We argue that the conventional SPT corresponds to a second-order quantum phase transition, whereas the unconventional SPT exhibits characteristics of a first-order quantum phase transition. Despite the model's non-$\mathcal{PT}$-symmetry, we observe a fully real energy spectrum under certain conditions.

Analysis of photon and atom populations, as well as fidelity measurements, indicates that the unconventional SPT lacks quantum criticality, characterized by the absence of asymptotic behavior and a sharp drop in fidelity at a single parameter value. In contrast, the conventional SPT exhibits quantum criticality, confirmed by critical exponents extracted from fidelity susceptibility scaling analyses, showing agreement with the single Rabi model's universality class.

Recent advancements in circuit-QED and ion trap experimental techniques offer substantial potential to implement and verify these identified SPTs in high-performance, state-of-the-art architectures. Moreover, these capabilities can be extended to investigate more complex coupled light-matter systems that currently surpass the capabilities of classical computational simulations.

We acknowledge the support from NSFC under Grant No. 12174346.

\end{document}